\documentclass[conference]{IEEEtran}

\IEEEoverridecommandlockouts
\usepackage{graphicx} % Required for inserting images
\usepackage{cite}
\usepackage{comment}
\usepackage{amsmath,amssymb,amsfonts}
\usepackage{algorithmic}
\usepackage[linesnumbered, ruled,vlined]{algorithm2e}
\usepackage{graphicx}
\usepackage{textcomp}
\usepackage{subcaption}
\usepackage{tikz}
\usepackage{soul}
\usetikzlibrary{positioning, shapes, arrows.meta, fit, decorations.text}

\usepackage[normalem]{ulem}
\usepackage{xcolor}
\usepackage{url}
\usetikzlibrary{calc, shapes, arrows, positioning}

\usepackage{xcolor}
\usepackage{authblk}
\def\BibTeX{{\rm B\kern-.05em{\sc i\kern-.025em b}\kern-.08em
    T\kern-.1667em\lower.7ex\hbox{E}\kern-.125emX}}

\usepackage{tikz}
\usetikzlibrary{positioning} % for "below=..." 语法
\usepackage{graphicx}         % for \resizebox
\newcommand{\normalize}{\normalsize}

\begin{document}

\title{A Decentralized Microservice Scheduling Approach Using Service Mesh in Cloud-Edge Systems\\
{\footnotesize \textsuperscript{}}
\thanks{}
}
\author[1]{Yangyang Wen}
\author[1]{Paul Townend}
\author[1]{Per-Olov Östberg}
\author[2]{Abel Souza}
\author[1]{Clément Courageux-Sudan}
\affil[1]{Umeå University, Umeå, Sweden \authorcr  {\{\emph{yangyang.wen, paul.townend, per-olov.ostberg, clement.courageux-sudan}\} @ umu.se}\vspace{1.5ex}}
\affil[2]{UC Santa Cruz, CA, USA \authorcr {absouza@ucsc.edu} \vspace{-2ex}} 

\maketitle

\begin{abstract}
As microservice-based systems scale across the cloud-edge continuum, traditional centralized scheduling mechanisms increasingly struggle with latency, coordination overhead, and fault tolerance. This paper presents a new architectural direction: leveraging service mesh sidecar proxies as decentralized, in-situ schedulers to enable scalable, low-latency coordination in large-scale, cloud-native environments. We propose embedding lightweight, autonomous scheduling logic into each sidecar, allowing scheduling decisions to be made locally without centralized control. This approach leverages the growing maturity of service mesh infrastructures, which support programmable distributed traffic management. We describe the design of such an architecture and present initial results demonstrating its scalability potential in terms of response time and latency under varying request rates. Rather than delivering a finalized scheduling algorithm, this paper presents a system-level architectural direction and preliminary evidence to support its scalability potential.

\end{abstract}


\section{Introduction}
Modern distributed systems are undergoing a paradigm shift, extending beyond centralized data centers into a heterogeneous cloud-to-edge continuum~\cite{patel2024}. With the rapid growth of federated cloud infrastructures and the proliferation of IoT devices, computation, and data are growing in scale and becoming increasingly decentralized. This evolution introduces new challenges for coordination, as traditional centralized models struggle to maintain efficiency in diverse high-load environments~\cite{Khallouli2022}. The scale and geographical distribution of these systems require new approaches to system management, particularly in areas such as scheduling and resource allocation. These trends are also reflected in the rise of stateless, on-demand execution models such as serverless computing, which further challenge traditional centralized scheduling designs.

The microservice architecture has become the dominant model for modern distributed systems \cite{b18}, breaking down applications into loosely coupled fine-grained services. In particular, microservices enable the replication of services for better scalability. However, this decoupling necessitates numerous network communications between services located on different machines. In this context, finding the best route between the service replicas to execute a request and meet performance objectives is a significant challenge.

A traditional approach to optimizing the performance of tasks executed within cloud-edge infrastructures relies on schedulers with global knowledge of the system. However, on scale and within dynamic environments, centralized models face inherent limitations. The increasing amount of requests to be scheduled makes schedulers a bottleneck, reducing the performance of applications\cite{senjab2023survey}.%{ADDCITATION}

The service mesh, typically implemented with Istio~\cite{istio} or Linkerd~\cite{Linkerd}, offers a programmable networking layer by deploying sidecar proxies alongside each service instance~\cite{b20}. These sidecars intercept and manage service traffic for better observability, security, and traffic control in a decentralized manner without modifying the application code. In particular, current service mesh solutions offer load balancing between microservice replicas for scalable scheduling of requests~\cite{b20}.

Inspired by this architecture, we investigate whether the network management capability of the service mesh can be extended beyond the classic load balancing algorithms and serve as autonomous in situ schedulers for scalable, low-latency request scheduling at the edge. Specifically, we explore a service scheduling approach centered around service mesh sidecars to make autonomous and local scheduling decisions. This design reduces coordination overhead and eliminates bottlenecks for significantly increased scalability and fault tolerance. To evaluate the proposed service-mesh-based scheduling framework, we instantiate it with a carbon-aware scheduling policy as a representative use case. However, the framework is designed to be agnostic to specific optimization goals and can support a wide range of scheduling strategies.

Prior research has explored decentralized scheduling and application-specific goals in isolation, but their combined impact on scalability in large-scale microservice systems remains underexplored~\cite{b5,b12}. Moreover, the impact of decentralized scheduling on scalability, particularly under high-load conditions, has not been systematically addressed.

This paper addresses: 1) To what extent can service mesh-based decentralized scheduling reduce coordination overhead and improve latency, particularly under increasing request concurrency in cloud-edge environments?; 2) How does the proposed sidecar-level decentralized scheduling architecture compare to a centralized scheduler in terms of makespan and stability under varying request rates?; and 3) What are the trade-offs introduced by decentralization in terms of performance consistency and system complexity?

The key contributions of this paper are as follows:

\begin{enumerate}
    \item We propose a novel decentralized scheduling architecture that leverages service mesh sidecar proxies to embed autonomous scheduling logic alongside microservices. This deployment model eliminates centralized coordination overhead while remaining compatible with modern cloud-native platforms.
    \item We design and implement a simulation-based experimental environment using SimGrid, which models real-world network latency, communication delays, execution times, and energy consumption across a network of multiple hosts. This platform enables a systematic comparison of the proposed decentralized scheduler with a MILP-based (Mixed Integer Linear Programming) centralized approach inspired by Casper~\cite{b0}, a representative centralized scheduler.
    \item We evaluate the system's scalability by comparing centralized and decentralized scheduling under varying workloads. The results demonstrate that the decentralized approach, enabled by the service mesh architecture, achieves a significantly lower makespan under high load, with minimal performance overhead.
\end{enumerate}

\section{Related Work}

\subsection{Centralized and Decentralized Scheduling}

A broad body of work has explored centralized and decentralized scheduling for improved scalability and responsiveness in distributed systems. Centralized schedulers, such as those used in early cloud and cluster environments, often suffer from scalability bottlenecks and single points of failure, particularly under high load or heterogeneous conditions. Omega~\cite{omega} proposes a shared-state, multi-scheduler architecture to mitigate contention among multiple scheduling components. Sparrow~\cite{sparrow} introduces a distributed, sampling-based scheduler aimed at low-latency task placement. Ray~\cite{ray} presents a distributed runtime for AI workloads with an actor-based model, enabling dynamic, locality-aware task execution.

%\add{
In centralized scheduling, global coordination relies on aggregating full system state into a centralized controller to perform global optimization, leading to scalability bottlenecks. In contrast, decentralized scheduling leverages lightweight coordination protocols (e.g., gossip, local state exchange) to achieve eventual consistency, enabling scalable and fault-tolerant scheduling decisions in large-scale cloud-edge environments.
%}

Recent research has increasingly focused on decentralized approaches, distributing scheduling decisions closer to the execution layer. Such strategies reduce coordination overhead and improve resilience under dynamic and large-scale conditions. Notably, Polaris~\cite{polaris} explores service mesh–compatible scheduling for cloud-native applications, while Hydra~\cite{hydra} avoids the bottleneck of centralized coordination by building a peer-to-peer overlay network between nodes, allowing each node to act as both a computing resource and undertake orchestration tasks.

\subsection{Scheduling for Serverless and Microservices}

The rise of Function-as-a-Service (FaaS) platforms has further emphasized the need for scalable, stateless, and dynamic scheduling mechanisms~\cite{Zuk2020, zhao2024serverless}. Statelessness, fine-grained execution, and elasticity in these environments present unique scheduling demands that challenge traditional centralized designs. Systems like Hydra~\cite{hydra} demonstrate the potential of decentralized control in such settings; however, these approaches overlook the challenges of scheduling at the microservice granularity, where long-running services, interdependencies, and locality awareness play a more critical role. Microservice-based environments introduce similar challenges at a finer granularity. Dyme~\cite{b13} and RESCAPE~\cite{b14} illustrate how service granularity can improve responsiveness. However, these works rely on centralized schedulers and do not address scalability concerns.

Scheduling algorithms in cloud computing have evolved to address various system constraints, such as sustainability. CASA~\cite{b10} and CASPER~\cite{b0} are centralized frameworks that optimize resource allocation to meet system-level goals, but they remain constrained by the limitations of global controllers in terms of scalability and fault tolerance.

Early efforts like GreenScale~\cite{b5} introduced edge computing with some degree of decentralization, but these systems still face challenges in achieving full scalability and microservice-level granularity. Recent works such as CarbonClipper~\cite{lechowicz2024carbonclipper} and Caspian~\cite{b3} continue to focus on centralized orchestration, which may not fully address the growing needs of modern cloud-edge systems.

\subsection{Classification Criteria}
%\add{
To compare existing scheduling and autoscaling approaches, we classify them based on three key dimensions: decentralization, scalability, and system overhead.
%}

\subsubsection{Decentralization}
We define centralized systems as those where a global controller makes all scheduling and scaling decisions.
\begin{itemize}
    \item Partial decentralized:
It refers to architectures where limited local autonomy exists under global policy control.
    \item Semi-decentralization:
It describes hierarchical systems where multiple local controllers handle fine-grained decisions while a global scheduler coordinates high-level resource allocation. 
    \item Fully decentralized: Fully decentralized systems distribute decision-making entirely among local agents with no global coordinator.
\end{itemize}

\subsubsection{Scalability}
We evaluate scalability based on the system's ability to handle increasing cluster sizes while maintaining performance and stability. Systems supporting only small-scale clusters (e.g., up to 8 nodes) are classified as \textit{Low}. Those successfully evaluated on tens of nodes (e.g., up to 32-50 nodes) are categorized as \textit{Moderate}. Systems capable of operating on hundreds or more nodes fall into the \textit{High} category.
\subsubsection{System Overhead}
System overhead reflects the computational and communication cost incurred during scheduling and auto scaling decisions at runtime. \textit{Low} overhead systems rely on static rules or threshold-based policies with minimal runtime computation. \textit{Moderate} overhead systems employ lightweight runtime heuristics, predictive models, or partial coordination among components. \textit{High} overhead systems involve complex runtime optimization (e.g., local search, reinforcement learning, or multi-objective trade-offs) and require extensive global state collection and coordination.

%\add{
Table~\ref{tab:comparisonSOTA} summarizes a number of existing scheduling approaches based on decentralization, scalability, and system overhead. \emph{Partial decentralized} refers to approaches where local nodes maintain limited autonomy under global control, while \emph{semi-decentralization} indicates hierarchical models combining global coordination with local schedulers. Scalability is classified based on the system's capability to support larger clusters while maintaining performance (e.g., from tens to hundreds of nodes). System overhead captures the computational and communication costs introduced by scheduling decisions, including runtime optimization complexity and control plane coordination.
%}

\begin{table}[h!]
\centering
    \caption{Comparison of existing Scheduling approaches}
    \label{tab:comparisonSOTA}
    \resizebox{\linewidth}{!}{
        \begin{tabular}{|l|c|c|c|}
        \hline
        \textbf{Approach} & \textbf{Decentralization} & \textbf{Scalability} & \textbf{System Overhead} \\ \hline
        CASA~\cite{b10} & Centralized & Moderate & High \\ \hline
        CASPER~\cite{b0} & Centralized & Moderate & Moderate \\ \hline
        GreenScale~\cite{b5} & Partial & Moderate & Moderate \\ \hline
        Hydra~\cite{hydra} & Fully Decentralized & High & Moderate \\ \hline
        Polaris~\cite{polaris} & Semi-Decentralized & High & High \\ \hline
        \textbf{This work} & \textbf{Fully Decentralized} & \textbf{High} & \textbf{Moderate} \\ \hline
        \end{tabular}
    }
\end{table}

The classification is based on the architecture of the system, the evaluation scale, and the complexity of the scheduling algorithm described in the original papers and follows a consistent interpretation across all compared methods.

Our approach leverages a sidecar-based fully decentralized scheduling framework, where each pod performs local decision-making using lightweight greedy algorithms based on real-time local observations. This design eliminates centralized coordination overhead, enabling high scalability, while the runtime cost of sidecars leads to moderate system overhead compared to purely reactive or centralized models.

\section{System Model and Assumptions} \label{AA}
%The following sections describe the system assumptions, \subsection{System Architecture}
\subsection{System Architecture}
The system consists of multiple geo-distributed regions, each with a cluster of compute nodes hosting stateless microservices. Services are pre-deployed in one or more regions, enabling redundancy and low-latency access.

Each service can execute a fixed number of requests in parallel using a thread pool. Service instances are paired with sidecars, responsible for the decentralized routing of requests between service replicas.

The system maintains an eventually consistent distributed metadata store to support decentralized scheduling. This shared control layer provides dynamic state information, such as resource utilization (e.g., CPU and memory, updated every few seconds), request queue statistics (which may fluctuate at millisecond granularity), and other relevant performance metrics (e.g., network load, latency). Propagation protocols, for instance, based on gossip can ensure the propagation of updates at scale without strong consistency but are out of the scope of this work.

Request handling is modeled as a queue-based mechanism: each service instance buffers incoming requests and executes them in a first-in, first-out (FIFO) manner to preserve arrival order. Service execution follows an actor-based concurrency model. This model enables parallel and distributed processing.

\subsection{Decentralized Scheduling Mechanism}
\begin{figure}[htbp]
\centering
\includegraphics[width=0.9\linewidth]{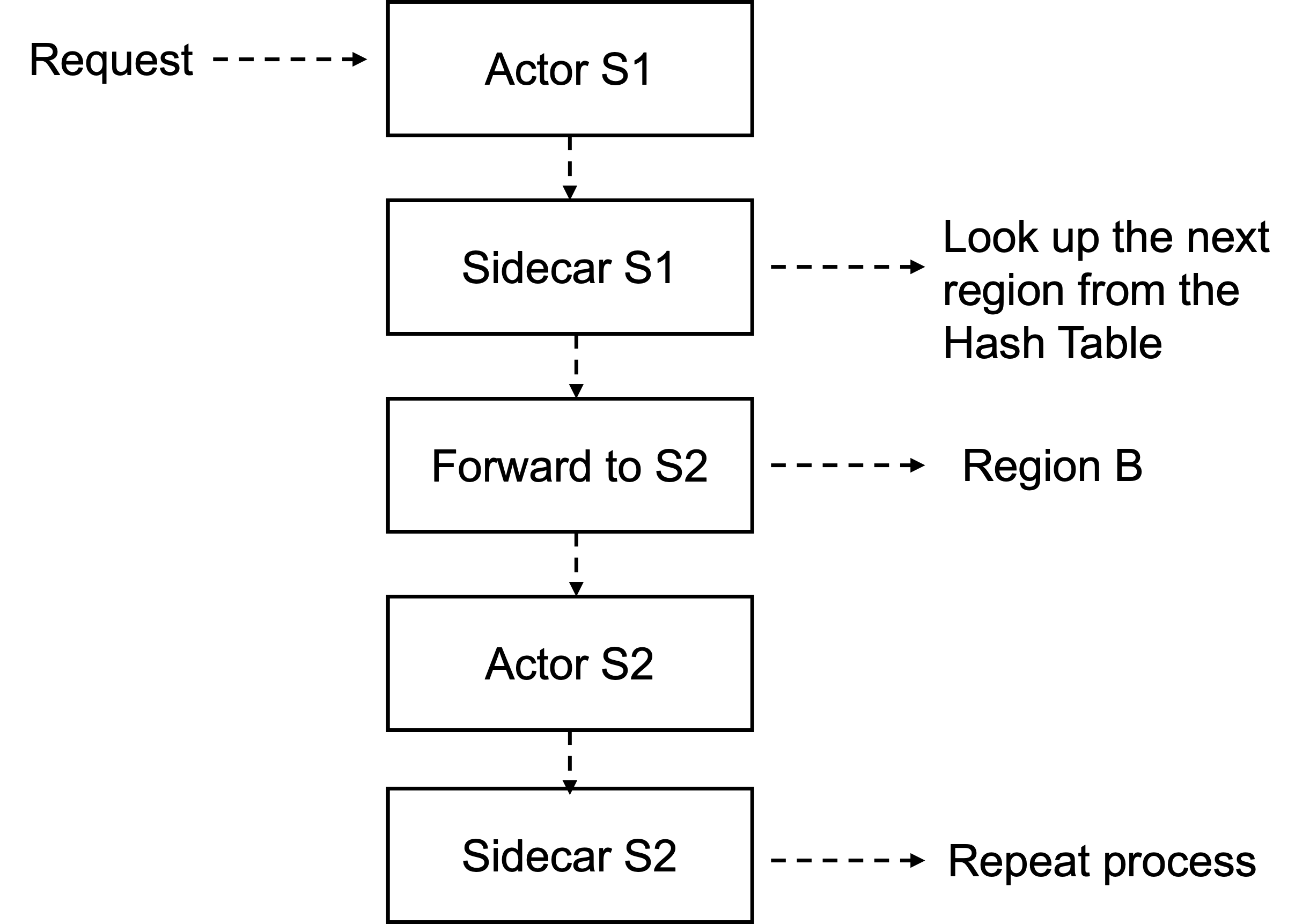}
\caption{Chain-based Execution Model: Hop-by-hop decentralized scheduling driven by sidecar agents using locally cached global metadata.}
\label{fig:chain-based}
\end{figure}

Scheduling follows a hop-by-hop, sidecar-driven approach, as shown in figure \ref{fig:chain-based}. Each application is represented as a service chain:

\[
SC = [S_1, S_2, ..., S_n]
\]

As soon as service $S_{i-1}$ finishes execution, its co-located sidecar selects an eligible host based on the scheduling objectives to execute $S_i$. The decision is taken based on the local cached view of the global metadata.

This strategy allows adaptive, decentralized scheduling; each sidecar makes decisions independently, improving fault tolerance and scalability. The architecture aligns with the Function-as-a-Service (FaaS) paradigm, where stateless, ephemeral (i.e., short-lived, and on-demand) services can scale elastically. Frequently used services remain warm, while others are gradually scaled down to preserve resources.

\subsection{Execution and Coordination Model}
Figure~\ref{fig:chain-based} illustrates the system’s architecture and the coordination mechanism. Key characteristics of our approach include: (1) Autonomous Scheduling: Each sidecar performs routing decisions independently using the local state. (2) Scalability: Stateless services scale dynamically based on observed demand. (3) Cold-start trade-offs: Cold starts are mitigated by retaining warm replicas in frequently active regions.

\paragraph{Assumptions and Discussion}
We assume:
(1) Services are stateless and replicated.
(2) Execution is modeled using an actor paradigm.
(3) Scheduling is handled by co-located sidecar agents.
(4) Metadata is shared via an eventually consistent global store.

\begin{figure}[h!]
\centering
\resizebox{\linewidth}{!}
{ % 限制图像宽度为当前栏宽
    \begin{tikzpicture}
    [
        region/.style={rectangle, draw=black, thick, minimum width=5.5cm, minimum height=4.5cm, label={[font=\Large]above:#1}},
        service/.style={rectangle, draw=blue!60, fill=blue!5, minimum width=1.6cm, minimum height=0.8cm, font=\Large},
        actor/.style={circle, draw=black!70, fill=gray!10, minimum size=0.7cm, font=\normalize},
        sidecar/.style={rectangle, draw=red!70, fill=red!10, minimum width=0.6cm, minimum height=0.5cm, font=\normalize},
        hash/.style={rectangle, draw=green!60!black, fill=green!5, thick, minimum width=6cm, minimum height=0.8cm, font=\Large},
        arrow/.style={-Latex, thick}
    ]
    
    % Region A
    \node[region=Region A, minimum width=7.8cm, minimum height=4.5cm] (regionA) at (0,0) {};
    \node[service, minimum width=3.3cm] (s1a) at (-1.9,1.2) {Service S1};
    \node[actor] (a1a) at (-2.8,0.2) {Actor11};
    \node[actor] (a2a) at (-1.0,0.2) {Actor12};
    \node[sidecar] (s1a1) [below=0.1cm of a1a] {Sidecar11};
    \node[sidecar] (s1a2) [below=0.1cm of a2a] {Sidecar12};
    
    \node[service, minimum width=3.3cm] (s2a) at (1.9,1.2) {Service S2};
    \node[actor] (a3a) at (1.0,0.2) {Actor21};
    \node[actor] (a4a) at (2.7,0.2) {Actor22};
    \node[sidecar] (s2a1) [below=0.1cm of a3a] {Sidecar21};
    \node[sidecar] (s2a2) [below=0.1cm of a4a] {Sidecar22};
    
    % Region B
    \node[region=Region B, minimum width=7.8cm, minimum height=4.5cm] (regionB) at (8,0) {};
    \node[service, minimum width=3.3cm] (s1b) at (6.0,1.2) {Service S1};
    \node[actor] (a1b) at (5.0,0.2) {Actor11};
    \node[actor] (a2b) at (6.8,0.2) {Actor12};
    \node[sidecar] (s1b1) [below=0.1cm of a1b] {Sidecar11};
    \node[sidecar] (s1b2) [below=0.1cm of a2b] {Sidecar12};
    
    \node[service, minimum width=3.3cm] (s3b) at (9.9,1.2) {Service S3};
    \node[actor] (a3b) at (8.9,0.2) {Actor31};
    \node[actor] (a4b) at (10.7,0.2) {Actor32};
    \node[sidecar] (s3b1) [below=0.1cm of a3b] {Sidecar31};
    \node[sidecar] (s3b2) [below=0.1cm of a4b] {Sidecar32};
    
    % Global Hash Table
    \node[hash] (hashtable) at (4,-2.5) {Eventually-consistent Distributed Metadata Store (e.g. resources, latency, utilization, etc.)}; %carbon etc.)};
    
    % Arrows from sidecars to hash table
    \foreach \x in {s1a1, s1a2, s2a1, s2a2, s1b1, s1b2, s3b1, s3b2} {
        \draw[arrow, dashed, gray] (\x) -- (hashtable);
    }
    
    \end{tikzpicture}
}
\caption{System Architecture: Geo-distributed regions host replicated services. Each actor instance is paired with a sidecar that performs local scheduling using metadata from a shared, eventually-consistent distributed metadata store.}
\label{fig:architecture}
\end{figure}
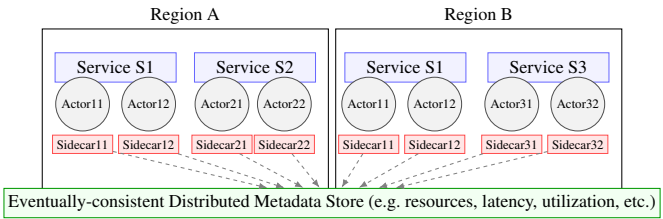

\section{Methodology}
In this section, we present our scheduling methodology for microservice requests, covering both centralized and decentralized approaches. We begin by defining a common scheduling objective, and then describe each approach in detail, including algorithms, architectural differences, and latency implications. Finally, we compare the computational complexity of the two scheduling strategies based on experimental runtime measurements. %empirical evaluations.

%These architectural choices are evaluated through simulation-based experiments, focusing on concurrency behavior, scalability, and response latency under various network and load conditions.

\subsection{Scheduling Objective}
%To enable a fair comparison between centralized and decentralized scheduling, we define a common objective. The goal in this work is to minimize the total carbon cost incurred by executing a microservice request across a service chain, while satisfying end-to-end latency constraints (QoS); however, our architecture supports any scheduling goal. 
To enable a fair comparison between centralized and decentralized scheduling, we define a common objective. The goal of this work is to minimize the overall resource utilization (or cost) incurred by executing a microservice request across a service chain while satisfying end-to-end latency constraints (QoS). Our architecture supports various scheduling goals beyond the one presented here.

%In this work, we demonstrate its effectiveness in a carbon-aware scheduling scenario. The objective is to minimize overall carbon emissions while ensuring latency remains within an acceptable threshold:

\[
\min_{r_1, ..., r_n} \sum_{i=1}^n \text{ResourceCost}(S_i, r_i)
\quad \text{s.t. } \text{Latency}(SC) \leq L_{\max},
\]

Where $\text{ResourceCost}(S_i, r_i)$ denotes the cost associated with executing service $S_i$ in region $r_i$, considering factors such as energy, computational, and resource usage, while $\text{Latency}(SC)$ is the end-to-end delay through the service chain. This formulation serves as a baseline for this work, but the framework can support a variety of alternative scheduling objectives.

%This formulation, the baseline for this work, is one of many possible optimization formulations; the proposed framework can support alternative scheduling objectives.

\subsection{Architectural Assumptions}

We model each microservice request as a chain of services, executed in sequence across a distributed infrastructure. Services are deployed as independent actor instances, each paired with a sidecar agent responsible for scheduling decisions. Asynchronous communication in the system is implemented using SimGrid, and the system architecture mimics some aspects of a service mesh, enabling communication between services.

Each sidecar maintains a metadata store for real-time metrics including:
\begin{itemize}
\item Resource usage across different regions,
%\item Carbon intensity in different regions,
\item Latency estimates to upstream services,
\item Resource availability per region.
\end{itemize}

These metrics are the basis for making scheduling decisions.

\subsection{Centralized Scheduler } %: Algorithms and Limitations}

In centralized scheduling architectures (e.g., Casper~\cite{b0}), a global controller computes the full execution path for each service chain request. Optimization is framed as a MILP problem, subject to resource and latency constraints:

\begin{itemize}
    \item A global scheduler selects hosts for all services in a request's chain.
    \item The MILP objective minimizes total emissions while satisfying CPU and network latency constraints.
    \item The availability of hosts is verified prior to optimization; the system fails if no valid mapping exists.
\end{itemize}
\begin{algorithm}
\caption{Centralized Scheduling with MILP Optimization}
\label{alg1}
\KwIn{Request $r$ with service chain $SC = [S_1, ..., S_n]$}
\KwOut{Request completion or failure due to constraints}

%\tcc{Step 1: 获取所有可用主机}
\For{each service $S_i$ in $SC$}{
    Find available hosts for $S_i$\;
    \If{no hosts available}{
        Return failure\;
    }
}

%\tcc{Step 2: 使用MILP求解器进行优化}
Define MILP solver\\
Define host selection variables $x_{i, h}$ for each service $S_i$ and host $h$\\
Set objective: Minimize carbon emission\\
Add constraints for CPU and latency\\

%\tcc{Step 3: 求解MILP并选择最优主机}
Solve MILP\;
\If{no optimal solution}{
    Return failure\;
}
Select optimal hosts for each service\;

%\tcc{Step 4: 生成执行路径}
For each selected host, retrieve corresponding mailbox\;
Calculate total latency between services\;
Return execution path\;

%\tcc{Step 5: 转发请求}
Send request to first mailbox\;
\end{algorithm}

We implement a customized MILP-based central scheduler (as shown in Algorithm \ref{alg1}) based on the formulation of Casper~\cite{b0}; once the MILP problem is solved, the optimal execution path is obtained from the set of constraints, and the request can be forwarded to the first service of the chain. All service instances are pre-provisioned and managed centrally, leading to potential bottlenecks and reduced elasticity under high load.

\subsection{Decentralized Scheduler}
\begin{algorithm}
\caption{Decentralized Scheduling via Sidecars}
\label{alg:decentralized-scheduling}
\KwIn{Request $r$ with service chain $SC = [S_1, ..., S_n]$}
\KwOut{Request completed or dropped}

Initialize $i \leftarrow 1$, currentRequest $\leftarrow r$\;

\While{$i \leq n$}{
    $S_i \leftarrow$ next service in chain\;
    Actor-sidecar at $S_i$ queries local metadata store: \\
    \Indp
        \quad Get system metric data (e.g., resource availability, latency, bandwidth, etc.)\;
    \Indm
    \If{no feasible region for $S_i$}{
        Drop request;\\ Return failure\;
    }
    Select execution region $r_i$ via local scoring rule (e.g., resource cost + latency penalty)\;
    
    Send request to actor replica of $S_i$ in region $r_i$\;
    
    Wait for $S_i$ execution to complete\;    
    $i \leftarrow i + 1$\;
}
Return success\;
\end{algorithm}
Our decentralized scheduler (as seen in Algorithm \ref{alg:decentralized-scheduling}) distributes control across services, enabling local decisions using sidecar agents. Each actor-sidecar pair selects an execution region by filtering candidates that meet both latency and resource availability constraints, before selecting the region with the best performance metrics (e.g., resource utilization and latency).

This model leverages the Actor Model~\cite{camilleri2022actor} for scalability and fault isolation. Communication is asynchronous, with localized scheduling introducing sidecar overhead at each hop.

\subsection{Modeling Scheduling Latency}

In addition to computational cost, the end-to-end request latency exhibits structural differences between centralized and decentralized scheduling. Under centralized scheduling, latency includes a one-time global scheduling delay followed by alternating transmission and processing stages along the service chain. In contrast, decentralized scheduling introduces a sidecar latency for each hop due to local executions in the sidecar, trading centralized coordination for per-hop overhead.

Given a service chain $S = \{S_1, S_2, \dots, S_n\}$, the total latency under centralized scheduling is:

\begin{equation}
L_{\text{central}} = L_{\text{sched}} + \sum_{i=1}^{n-1} \left( \text{TWL}_i + L_{\text{proc}}^{(i)} \right)
+ L_{\text{proc}}^{(n)}
\end{equation}

where $L_{\text{sched}}$ is the centralized scheduling latency, $\text{TWL}_i$ is the transmission latency before service $S_i$, and $L_{\text{proc}}^{(i)}$ is the processing latency of service $S_i$.

Under decentralized scheduling, the total latency is:

\begin{equation}
L_{\text{decentral}} = \sum_{i=1}^{n-1} \left( \text{TWL}_i + L_{\text{proc}}^{(i)} + L_{\text{sidecar}}^{(i)} \right) + \text{TWL}_n + L_{\text{proc}}^{(n)}
\end{equation}

Here, $L_{\text{sidecar}}^{(i)}$ represents the sidecar scheduling delay introduced after each intermediate service hop. The final hop omits sidecar latency, assuming no further decision is needed post-processing.

\subsection{Complexity comparison of scheduling algorithms}
To better understand the scalability of each scheduling approach, we analyze their computational complexity in terms of service chain length and replica count. We implement both in C++ and profile their execution time; initial empirical results are presented in Section~\ref{sec:results-complexity}.

\section{Implementation using SimGrid}
To realistically evaluate our scheduling strategies, we implement the scheduling approaches using the SimGrid simulator. This section details how we implement the two schedulers, the microservice applications and service mesh sidecars.

\subsection{Actor-Based Execution Model}

To simulate resource usage in a cloud-edge environment, we adopt an execution model implemented within the SimGrid simulation framework~\cite{CASANOVA2025103125}. This model captures the interplay between task concurrency and CPU core allocation, representing realistic constraints of distributed resource scheduling.
The simulation of a computational machine (host) consists of two primary components:
\begin{itemize}
    \item A physical CPU, configured with several cores, frequency, and energy profiles.
    \item A set of actors simulating the resource usage of processes (service handlers, sidecar) running on the CPU and using simulated network interfaces.
\end{itemize}

Our service mesh scheduling simulations are governed by the following. 

The key parameters governing our service mesh scheduling simulations are summarized in Table~\ref{tab:sim_parameters}.

\begin{table}[h!]
\centering
\caption{Simulation Parameters}
\begin{tabular}{|c|l|}
\hline
\textbf{Symbol} & \textbf{Description} \\
\hline
$A$ & Number of parallel actors per host per service replica \\
$C$ & Number of physical CPU cores on each host \\
$\lambda$ & Request arrival rate (requests per time unit) \\
$T_{\text{comp}}$ & Fixed computation time per request (ideal, no contention) \\
\hline
\end{tabular}
\label{tab:sim_parameters}
\end{table}

\subsection{Execution Scenarios and Resource Contention}

We consider two configurations based on the relationship between $A$ and $C$:

\paragraph{Case 1: $C \geq A$ and no CPU bottleneck (Underloaded or Balanced System)} 

In this case, sufficient CPU cores are there to serve all requests instantly. When a request arrives on a machine:
\begin{itemize}
    \item It is distributed to one of the available actors.
    \item That actor makes exclusive use of a CPU core to run the request with a computation time $T_{\text{comp}}$, and a waiting time is negligible unless the arrival rate $\lambda$ exceeds the processing capability of the CPU.
\end{itemize}

The maximum concurrency is thus limited by $A$:
\[
N_{\text{concurrent}} = \min(\lambda, A)
\]
The system's CPU utilization is bounded by:
\[
U_{\text{CPU}} = \frac{\min(\lambda, A) \cdot T_{\text{comp}}}{C \cdot T_{\text{comp}}} = \frac{\min(\lambda, A)}{C}
\]

\paragraph{Case 2: $C < A$ (Overloaded System)}

In this setting, the number of actors exceeds the number of CPU cores, leading to potential oversubscription. SimGrid models CPU contention through time-sharing. Each actor receives a fraction of the available compute capacity, resulting in extended execution times and increased latency:
\begin{itemize}
    \item When all cores are busy, actors share CPU time.
    \item Each active actor experiences extended computation time due to time-slicing and CPU contention.
    \item Let $n_{\text{active}}$ denote the number of simultaneously executing actors; if $n_{\text{active}} > C$, then each actor receives $\frac{1}{n_{\text{active}}}$ of a core.
\end{itemize}

The effective computation time per request becomes:
\[
T'_{\text{comp}} = T_{\text{comp}} \cdot \frac{n_{\text{active}}}{C}
\]

This slowdown affects both request latency and overall makespan. If the request arrival rate $\lambda$ exceeds $A$, then queueing occurs, and waiting time contributes to total delay:
\[
T_{\text{total}} = T_{\text{wait}} + T'_{\text{comp}}
\]

\subsection{Concurrency and Scalability Insights}
Concurrency is a foundational concept in cloud computing, enabling efficient task execution and resource utilization through parallel processing~\cite{armbrust2009above, dean2008mapreduce}. In our model, we capture hard parallelism by assigning multiple CPU cores per host and implementing soft concurrency through the number of available actors. This separation allows us to explore how system throughput and latency respond to both physical and logical concurrency constraints:

%This actor-CPU model captures the essence of parallel execution constraints in cloud computing. We distinguish between:
\begin{itemize}
    \item Hard parallelism: determined by the number of CPU cores $C$.
    \item Soft concurrency: determined by the number of actors $A$ that can handle simultaneous requests.
\end{itemize}

By tuning $A$ and $C$, we emulate both saturation scenarios and fully parallelizable workloads. For instance, if $\lambda \leq C$, requests are processed without contention; if $C < \lambda \leq A$, concurrent processing occurs with extended $T'_{\text{comp}}$; if $\lambda > A$, queuing becomes inevitable.

This abstraction supports a realistic evaluation of cloud scheduling strategies under dynamic workloads and hardware constraints.

\subsection{Evaluation Architecture for Concurrency and Scheduling Comparison}
To evaluate scheduling overhead and scalability in isolation, we adopt the default cluster topology recommended by SimGrid\cite{simgrid2025platformexamples}, which in our case consists of 1100 hosts connected via homogeneous network links and grouped into 10 logical regions. While the framework can support more realistic energy or latency-aware models, we focus specifically on the scheduling layer and abstract away detailed bandwidth or geographic latency differences. Historical region-level heterogeneity (e.g., carbon intensity) is introduced to support a comparative evaluation of decentralized behavior, with our scope centered on scheduling logic rather than the complete end-to-end service execution process (from request submission to response delivery).

To ensure a fair comparison between centralized and decentralized scheduling approaches, both of them are evaluated on the same network topology. The infrastructure includes 1100 hosts, each with 24 CPU cores, an Ethernet network interface, and an energy profile as summarized in Table~\ref{tab:simgrid_config}.

\begin{table}[htbp]
\centering
\caption{Key SimGrid Host Configuration Parameters}
\label{tab:simgrid_config}
\begin{tabular}{|l|l|}
\hline
\textbf{Parameter} & \textbf{Value} \\ \hline
Number of Hosts & 1100 \\ \hline
CPU Cores per Host & 24 \\ \hline
CPU Speed & 10 GFLOPS \\ \hline
Network Bandwidth & 125 MBps \\ \hline
Network Latency & 50 $\mu$s \\ \hline
Backbone Bandwidth & 2.25 GBps \\ \hline
Backbone Latency & 500 $\mu$s \\ \hline
Power (Off State) & 10 W \\ \hline
Power (Idle: 0\% usage) & 20 W \\ \hline
Power (Max: 100\% usage) & 200 W \\ \hline
\end{tabular}
\end{table}

\section{Evaluation}\label{ES}
%\add{
In evaluating scalability, we focus primarily on the scheduler's decision latency growth under increasing request rates. While resource utilization reflects scheduling optimality, it is orthogonal to the system's scalability capabilities.
%}

This section presents the initial evaluation of our centralized and decentralized scheduling strategies using the SimGrid simulation framework. We compare their performance across varying workloads, focusing on request makespan and system scalability. The simulation is still under development and thus results are only indicative until comparatively evaluated against a real implementation.

\subsection{Simulation Environment and Experimental Setup}
%Experiments are conducted using SimGrid, a flow-level simulator for modeling cloud-edge infrastructures. 

Experiments are conducted using the SimGrid simulation framework~\cite{CASANOVA2025103125}. We chose SimGrid, a flow-level simulator because of its capability to accurately simulate the network and application performance of cloud-edge infrastructures, including microservices~\cite{courageux2021automated}.\footnote{A flow-level simulator models communication at the granularity of logical flows (e.g., data transfers between services), rather than simulating individual packets. This abstraction significantly improves simulation speed while maintaining sufficient accuracy for evaluating large-scale distributed systems.} We simulate a system of 1100 hosts distributed across 10 regions. Each service in the 10-stage microservice workflow has 50 replicas, evenly placed across the regions.

Requests are generated at rates ranging from 1 to 15,000 requests per second (rps). In the centralized setting, requests are routed through a global scheduler node; in the decentralized setting, decisions are made locally at each service node. Tables~\ref{tab:centralized_hosts} and~\ref{tab:decentralized_hosts} summarize the infrastructure allocation under both scheduling modes. Table \ref{tab:exp_setting} summarizes the distribution of servers, replicas, and system metrics across the regions, for both centralized and decentralized scheduling.

%the carbon intensity mapping across the regions, for both centralized and decentralized scheduling.

The primary metric is the end-to-end makespan per request, defined as the elapsed time from request generation to final response. It can capture the cumulative impact of scheduling decisions, network latency, and resource contention.

\begin{table}[htbp]
\centering
\caption{Hosts distribution for the centralized algorithm.}
\begin{tabular}{|c|c|}
\hline
\textbf{Region} & \textbf{Number of Hosts (Centralized)} \\ \hline
Region 1-9 & 109 hosts each (Total 981 hosts) \\ \hline
Region 10 & 117 hosts \\ \hline
Client & 1 host (randomly assigned region) \\ \hline
Central Scheduler & 1 host (randomly assigned region) \\ \hline 
\end{tabular}
\label{tab:centralized_hosts}
\end{table}

\begin{table}[htbp]
\centering
\caption{Hosts distribution for the decentralized algorithm.}
\begin{tabular}{|c|c|}
\hline
\textbf{Region} & \textbf{Number of Hosts (Decentralized)} \\ \hline
Region 1-9 & 109 hosts each (Total 981 hosts) \\ \hline
Region 10 & 118 hosts \\ \hline
Client & 1 host (randomly assigned region) \\ \hline
Server (Replica Execution) & 1099 hosts (randomly assigned region)\\ \hline
\end{tabular}
\label{tab:decentralized_hosts}
\end{table}

%\subsection{System Configuration Table}
%The following Table \ref{tab:exp_setting} summarizes the distribution of servers, replicas, and the carbon intensity mapping across the regions, for both centralized and decentralized scheduling:
\begin{table}[htbp]
\centering
\caption{System and Request Configuration}
\begin{tabular}{|l|l|}
\hline
\textbf{Item} & \textbf{Description} \\
\hline
Number of request types & 1 \\
Request length & 10 (involves services S1 to S10) \\
Number of service types & 10 \\
Replicas per service & 50 \\
Total number of services & 500 \\
Number of regions in system & 10 \\
Hosts per region & Around 100 \\
Total number of hosts & Over 1000 \\
Replica placement & Distributed across hosts in all regions \\
\hline
\end{tabular}
\label{tab:exp_setting}
\end{table}

\section{Results and Analysis}

\subsection{Empirical results}\label{sec:results-complexity}
\begin{figure}[h!]
\centering
    \begin{subfigure}{\linewidth}
      \includegraphics[trim=4cm 0.6cm 1cm 2.2cm, clip, width=1.0\linewidth]{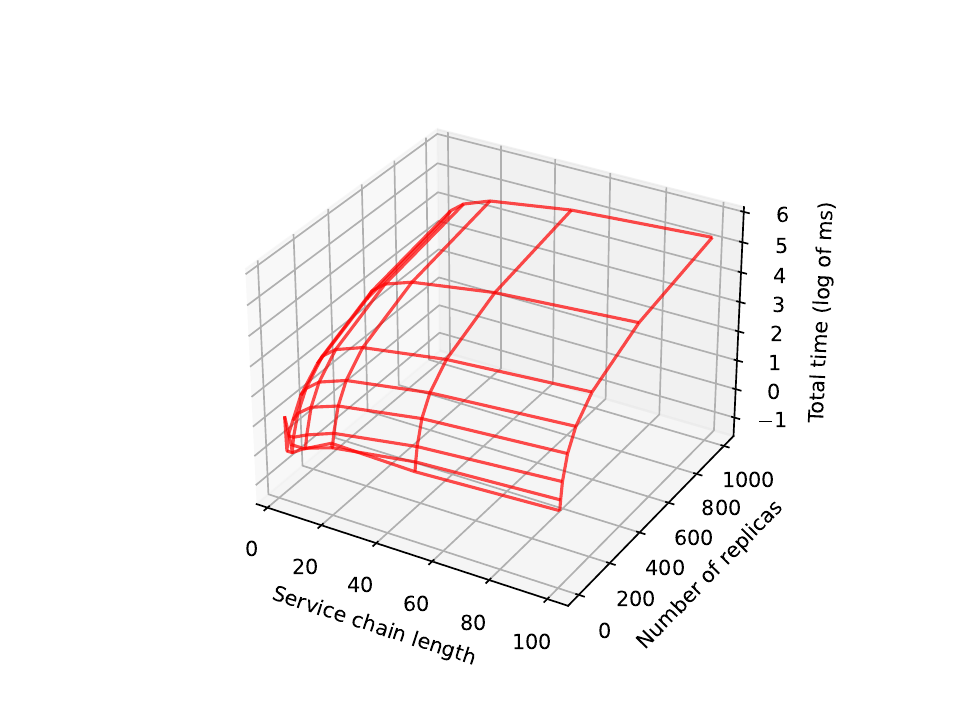}
      \caption{Centralized scheduling execution time}
      \label{fig:centralized-flops}
    \end{subfigure}
    \hfill
    \vspace{2.5cm}  % 调整这个值以控制间距
    \begin{subfigure}{\linewidth}
      \includegraphics[trim=4cm 0.6cm 1cm 2.2cm, width=1.0\linewidth]{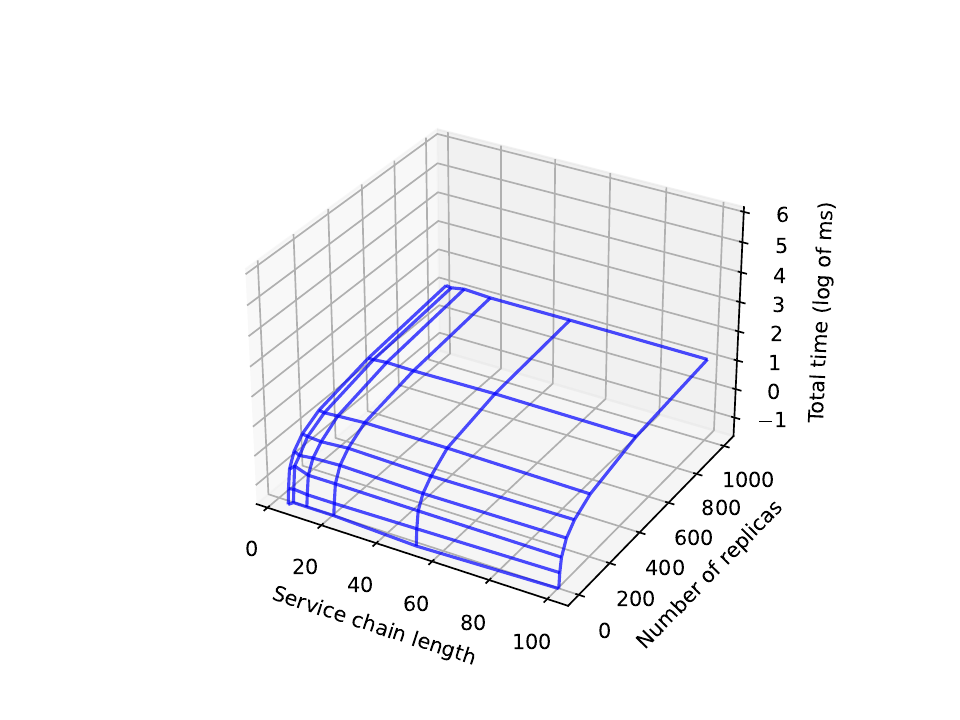}
      \caption{Decentralized scheduling execution time}
      \label{fig:decentralized-flops}
    \end{subfigure}
    \caption{3D surface plots of scheduler wallclock time as a function of service chain length and replica count. Centralized scheduling incurs steep growth in computational demand, while decentralized scheduling remains more scalable.}
    \label{fig:flops-3d}
\end{figure}

To quantify the runtime impact of centralized vs. decentralized scheduling, we measure the computational time taken for scheduling decisions.
We implement both centralized and decentralized algorithms in C++, and execute them under varying configurations. We use varying service chain lengths $\{3, 5, 10, 20, 50, 100\}$ and service replicas counts $\{5, 10, 20, 50, 100, 200, 500, 1000\}$. For each configuration, we record the average wall-clock time over five executions to reduce the impact of external interferences. Figure~\ref{fig:centralized-flops} depicts the duration of processing requests with the centralized approach and Figure~\ref{fig:decentralized-flops} for the decentralized approach.

Empirical observations match theoretical expectations: increasing the search space with longer chains and a higher number of replicas leads to longer execution times for centralized scheduling than for a decentralized approach dividing the search space.  

\subsection{Performance Comparison}

We perform an analysis of makespan for varying request rates using our simulator. The makespan is defined as the total time taken to execute a microservice request including the scheduling time, in seconds. The makespan was extracted from the simulations' output for both algorithms, and the average makespan was computed.% for each request rate.

\begin{figure}[htbp]
    \centering
    \includegraphics[width=1.0\linewidth]{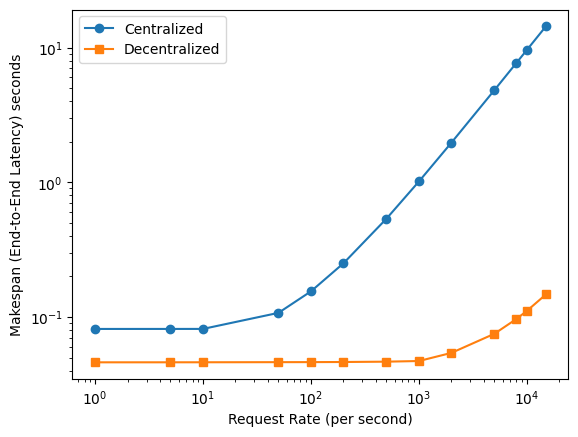}
    \caption{Indicative Simulation Results Showing Makespan Comparison between Centralized and Decentralized Scheduling Algorithms}
    \label{fig:final_results}
\end{figure}

Figure~\ref{fig:final_results} shows indicative simulation results of the end-to-end latency of request executions against the rate of requests sent to applications per second.

The graph indicates that the centralized algorithm exhibits lower makespan at lower request rates but suffers significant latency increases as load grows, due to the overhead of centralized decision-making. In contrast, the decentralized algorithm results indicate a more consistent performance as the system scales. These performance indications show a decentralized service mesh-based approach has promise, but our findings still need to be validated in a real hardware environment.

\section{Challenges in Decentralized Scheduling}

While our proposed architecture demonstrates the feasibility and scalability of decentralized scheduling via local sidecar agents, it also introduces several open challenges that warrant further exploration:

\paragraph{Load Balancing and Performance Consistency}
Local decision-making based on incomplete system metadata can lead to coordination blind spots—for example, multiple agents independently selecting the same underloaded node during bursty traffic, resulting in resource contention and performance degradation. Designing lightweight coordination or feedback mechanisms remains an open research direction.

\paragraph{Operational Complexity and Observability}
The shift from centralized to distributed control increases the complexity of monitoring, debugging, and managing system-wide behavior. Effective abstractions and observability tools must be developed to maintain transparency and debuggability without reintroducing centralized bottlenecks.

\paragraph{Policy Conflicts and Heuristic Divergence}
Autonomous sidecars may implement heterogeneous heuristics, especially in multi-tenant or cross-team environments. This lack of harmonization can lead to inefficient resource usage or conflicting scheduling decisions. Establishing interoperable coordination protocols or shared policy layers may help align local actions with global objectives.

\paragraph{Fault Tolerance and Failure Containment}
Without centralized oversight, local failures (e.g., misbehaving sidecars or overloaded replicas) may propagate more easily. Future designs must address how to isolate faults, introduce redundancy, or apply selective consensus without compromising the system's scalability.

These challenges define the research frontier for decentralized scheduling and highlight the necessary trade-offs between autonomy, coordination, and operational simplicity in distributed service infrastructures.

\section{Conclusion}

This work proposes a decentralized, actor-based scheduling architecture for stateless microservices deployed across geo-distributed regions; each service replica is paired with a lightweight sidecar agent that performs hop-by-hop scheduling based on locally synchronized metadata from a distributed hash table. This aims to eliminate centralized bottlenecks - enabling elastic scalability, improved responsiveness under dynamic workloads, and efficient decision-making based on local system metrics.

The approach aligns with the growing trend of Function-as-a-Service (FaaS) and microservices architectures, where statelessness and locality facilitate adaptive control.

Initial simulation-based results provide an indication that our decentralized approach (despite being based on a greedy path selection) outperforms centralized scheduling in terms of scalability and makespan, especially under high request rates. The results also indicate challenges related to balancing autonomy with global coordination.

Future work will begin with a comprehensive validation of this initial simulation-based work through implementation on a real Kubernetes-based cluster environment. We will use this implementation to also investigate additional optimization metrics, including cost, resource utilization, and latency.

Subsequent to this, we aim to focus on improving scheduling resilience in the presence of stale or partial metadata. Additionally, we will explore hybrid scheduling approaches that combine local autonomy with limited global coordination to address the challenges of failure isolation and coordination under bursty loads.

\section*{Acknowledgments}
This work was partially supported by the Wallenberg AI, Autonomous Systems and Software Program (WASP) funded by the Knut and Alice Wallenberg Foundation. This work is also supported by the European Commission through the Horizon Europe project SovereignEdge.Cognit under Grant Agreement 101092711.

%\balance
%\bibliographystyle{IEEEtranS}
%\bibliographystyle{IEEEtran}  % 或 ieeetr
%\bibliography{biblio}

%\bibliographystyle{ieeetr}
%\bibliography{biblio}

\end{document}